\documentclass{optica-article}

\journal{opticajournal} 

\articletype{Research Article}
\usepackage{lineno}
\usepackage{tabularray}
\UseTblrLibrary{diagbox}
\linenumbers 

\begin{document}
\nolinenumbers
\title{Frequency-aware optical coherence tomography image super-resolution via conditional generative adversarial neural network}

\author{Xueshen Li\authormark{1}, 
Zhenxing Dong\authormark{2},
Hongshan Liu\authormark{1},
Jennifer J. Kang-Mieler\authormark{1},
Yuye Ling\authormark{2},
and Yu Gan\authormark{1,*}}

\address{\authormark{1}Department of Biomedical Engineering, Stevens Institute of Technology}
\address{\authormark{2}Department of Electronic Engineering, Shanghai JiaoTong University}

\email{\authormark{*}ygan5@stevens.edu} 

\begin{abstract*} 
Optical coherence tomography (OCT) has stimulated a wide range of medical image-based diagnosis and treatment in fields such as cardiology and ophthalmology. Such applications can be further facilitated by deep learning-based super-resolution technology, which improves the capability of resolving morphological structures. However, existing deep learning-based method only focuses on spatial distribution and disregard frequency fidelity in image reconstruction, leading to a frequency bias. To overcome this limitation, we propose a frequency-aware super-resolution framework that integrates three critical frequency-based modules (i.e., frequency transformation, frequency skip connection, and frequency alignment) and frequency-based loss function into a conditional generative adversarial network (cGAN). We conducted a large-scale quantitative study from an existing coronary OCT dataset to demonstrate the superiority of our proposed framework over existing deep learning frameworks. In addition, we confirmed the generalizability of our framework by applying it to fish corneal images and rat retinal images, demonstrating its capability to super-resolve morphological details in eye imaging. 
\end{abstract*}

\section{Introduction}

Optical coherence tomography (OCT) is a non-invasive imaging modality that utilizes infrared interferometry to generate depth-resolved reflectivity profile in real-time \cite{YuyeLing.2023}. Over the last decades, OCT has stimulated a wide range of medical image-based diagnosis and treatment \cite{AlMujaini.2013, li2022structural, Liu.2023}. For example, in cardiology, OCT is considered a suitable coronary imaging modality to assess plaques to ensure successful stent deployment \cite{Lee.2013}. Meanwhile, in ophthalmology, OCT has become one of the prominent diagnostic tools for keratoconus \cite{Yang.2021}, glaucoma \cite{Kamalipour.2021}, age-related macular degeneration \cite{Scuteri.2019}, retinopathy \cite{Midena.2021}, and diabetic retinopathy and diabetic macular edemas \cite{Sikorski.2013} in identifying layers in both anterior and posterior segments of the eye. 

In both coronary imaging and eye imaging, high spatial resolutions from OCT, mostly spectral domain OCT (SDOCT), are crucial to facilitate the application to either identify endothelial cells or assess the thickness of corneal layers and retinal layers. However, such high resolution comes at the cost of demanding optical design and data transmission/storage. Improvement of resolution via upgrading light sources and other hardware designs is resource-intensive. The hardware improvement also suffers from jittering and motion artifacts caused by sparse sampling. On the contrary, software-based method could bypass the hardware upgrading issue and achieve high image quality through computational methods.


In the realm of algorithmic super-resolution (SR), various digital signal processing and image processing methods, have been developed to generate high-resolution (HR) OCT images from low-resolution (LR) OCT scanning which is undersampled in spectral or/and spatial domain. Conventionally, deconvolution \cite{Liu.2009, Hojjatoleslami.2013}, spectrum-shaping \cite{Chen.2017}, and spectral estimation \cite{Liu.2015} have been proposed to optimize OCT images. However, the SR performance is recently boosted by the introduction of deep learning (DL), especially the combination of convolution neural network (CNN) and generative adversarial network (GAN).  

Convolution neural network has been widely used for OCT image generation \cite{YingXu.2018, KaichengLiang.2020, QiangjiangHao.2020, Lichtenegger.2021b, Qiu.2021, Zhang.2021b, Li.2022, Ronneberger.2015b} to enhance the image quality and denoise speckles \cite{QiangjiangHao.2020, YuyeLing.2023}. In particular, CNN models such as multi-scale residual network (MSRN), residual dense network (RDN), residual dense Unet (RDU), and Residual Channel Attention Network
(RCAN) have been utilized and compared recently in generating SR OCT images \cite{Li.2022}. Moreover,
Conditional generative adversarial network (cGAN) has been incorporated in OCT SR research \cite{9252899, QiangjiangHao.2020, HongmingPan.2020, Das.2020}, featuring a discriminator design to examine the fidelity of the generated SR image during the training process, thus enhancing the capability of generating HR images. 

\begin{figure*}[t]
\centering
\includegraphics[width=\textwidth]{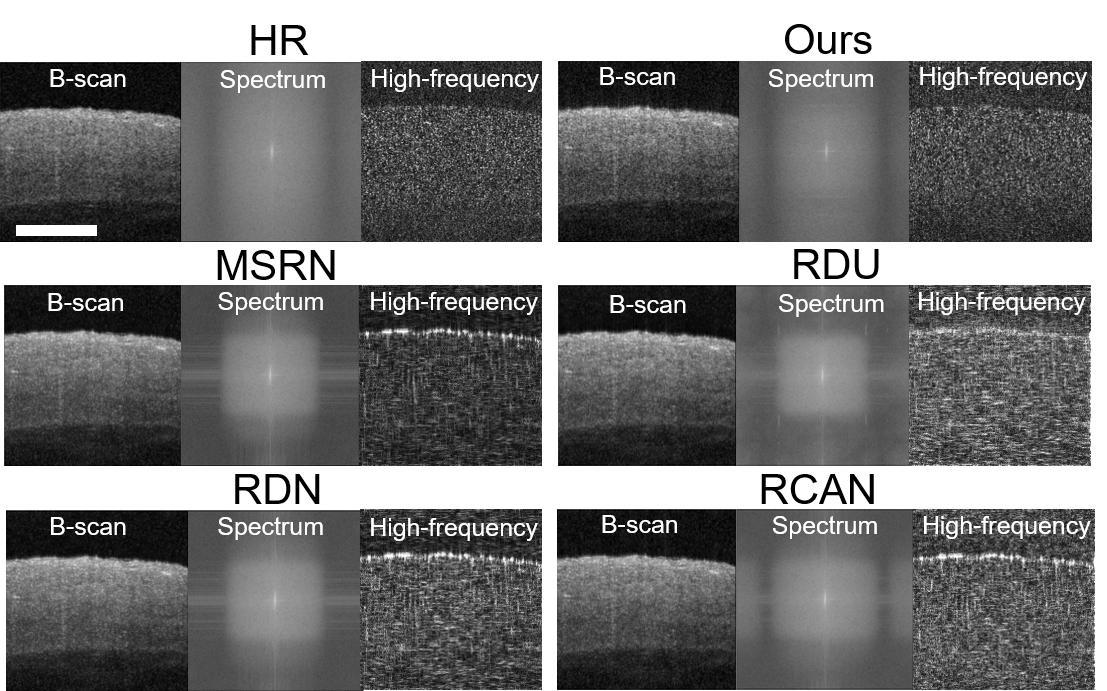}
\caption{Frequency domain gaps between the HR and the SR OCT images generated by four CNN implementations (MSRN, RDN, RDU, RCAN). The spectrum is generated by performing Fourier transform on the B-scan image. The high-frequency components of images are generated by performing an inverse Fourier transform on the high-frequency parts of the spectrum. Compared to the HR image, SR images generated by existing CNN algorithms are biased to a limited spectrum region towards low-frequency. Using ours CNN algorithm with frequency awareness (ours), the spectrum and high-frequency components of the SR OCT image are closer to that of the original image. The scale bar represents 500$\mu$m.}
\label{fig:freqCompare}
\end{figure*}

However, the current DL research in generating SR OCT images focuses solely on the spatial distribution of pixels in B-scans, without consideration of frequency information. The lack of frequency-awareness poses limitation on SR performance in two aspects. Firstly, from 1-D frequency perspective, SDOCT is physically measured in spectrum and reconstructed in spatial domain. Considering frequency information along axial direction would increase the fidelity of reconstruction. Secondly, 
from 2-D image processing perspective, current DL models exhibit spectral bias, which is a learning bias towards low-frequency components \cite{Rahaman.2019, 10.5555/3495724.3496356}. As shown in Fig \ref{fig:freqCompare}, DL algorithms induce frequency domain gaps in SR OCT images compared to the reference HR images, as they fail to resemble the high-frequency components, such as edges and textures of the coronary artery sample. High-frequency components preserve finer details that are beneficial for medical imaging \cite{9469469}. 
Therefore, a DL framework with frequency awareness is needed to reduce spectral bias and generate high-quality SR OCT images.

To this end, we develop a deep learning framework that is capable of super-resolving LR OCT images with frequency awareness. In this manuscript, we propose a DL framework for OCT image SR task with frequency awareness, which is capable of restoring high-frequency information via model design and optimized loss function. We perform extensive experiments on an existing human coronary dataset and quantitatively demonstrate that the proposed frequency-aware DL framework super-resolves OCT images with superior quality and less frequency bias. We also validate the spectral bias of the existing DL algorithms used in generating SR OCT images. Furthermore, we perform qualitative analysis to confirm that our framework is capable of generating SR OCT images for corneal imaging and retinal imaging.

\section{Methods}

\subsection{Overall framework}
\begin{figure*}[t]
\centering
\includegraphics[width=\textwidth]{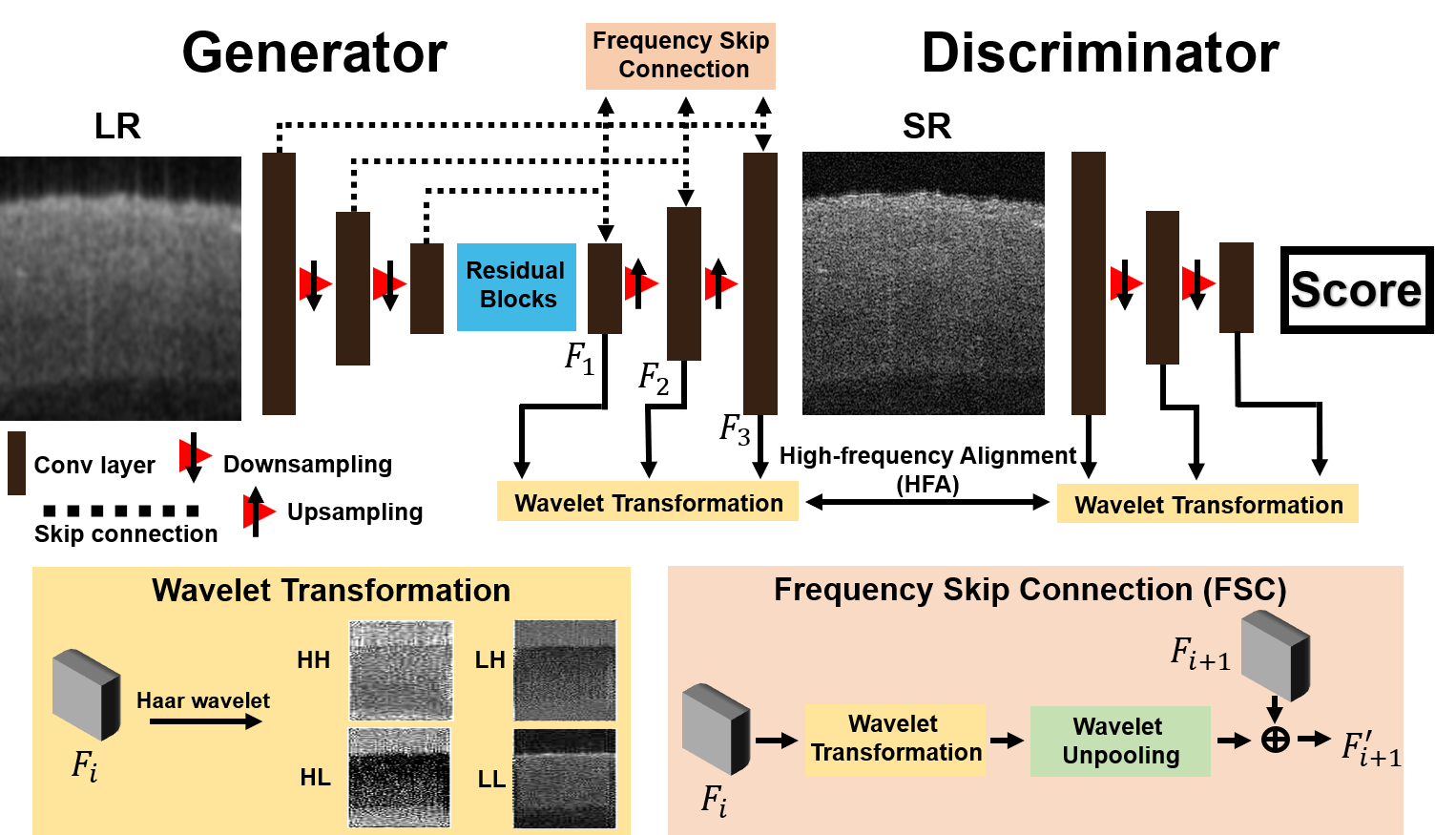}
\caption{The design of the proposed frequency-aware framework for OCT image super-resolution. The proposed model utilizes wavelet transformation, frequency skip connection, and high-frequency alignment to facilitate frequency information for super-resolving OCT images.}
\label{fig:struc}
\end{figure*}

The design of our frequency-aware framework is shown in Fig \ref{fig:struc}. Our framework consists of a generator ($G$) and a discriminator ($D$). Generator $G$ translates a LR image to a SR image. Discriminator $D$ classifies whether or not the generated image is realistic. Wavelet transformation is utilized to decompose feature maps $F_i$ into different frequency components; frequency skip connection (FSC) is used to prevent the loss of high-frequency information; high-frequency alignment (HFA) is used to guide the G for generating frequency information \cite{yang.2022}.
\subsection{Model design}
\subsubsection{Wavelet transformation} 
We adopt Haar wavelet for decomposing feature maps of the $i$-th layer $F_i$ into different components. Haar wavelet consists of two mirror operations: wavelet pooling and wavelet unpooling. The wavelet pooling converts images into the wavelet domain, and the wavelet unpooling inversely reconstructs frequency components into the spatial domain. During wavelet pooling, $F_i$ is convolved with four distinct filters: $LL^T$,  $LH^T$, $HL^T$, and $HH^T$, where $L$ and $H$ are low and high pass filters respectively ($L^T=\frac{1}{\sqrt{2}}[1,1]$, $H^T=\frac{1}{\sqrt{2}}[-1,1]$). The low pass filter ($LL^T$) provides general shapes and outlines in $F_i$; the high pass filters ($LH^T$, $HL^T$, $HH^T$) provide more fine details such as segments, edges, and contrasts. An illustration of wavelet transformation is shown in Fig \ref{fig:struc}. 

\subsubsection{Frequency skip connection} 
To prevent the loss of high-frequency information from $F_i$ to $F_{i+1}$, FSC is used in generator $G$. The FSC in G is defined as:
\begin{equation}
F^{'}_{i+1}=F_{i+1}+Unpooling(LL_{G}^{i}, LH_{G}^{i}, HL_{G}^{i}, HH_{G}^{i})
\end{equation}
After the frequency skip connection, feature map $F^{'}_{i+1}$ is calculated with the frequency information of $F^{'}_{i}$ is preserved through this process.

\subsubsection{High-frequency alignment} 
High-frequency alignment (HFA) provides $G$ with a self-supervised learning scheme using frequency information acquired in $D$. For $F_i$ in G, we acquire $LL_G^i$,  $LH_G^i$, $HL_G^i$, and $HH_G^i$. The combination of high-frequency components in $G$ is defined by: $HF_G^i=LH_G^i+HL_G^i+HH_G^i$. Similarly, high-frequency components in $D$ can be acquired by $HF_D^i=LH_D^i+HL_D^i+HH_D^i$. The $HF_D^i$ can be used as the self-supervision constraint to train $G$.

\subsection{Loss function}
In the proposed frequency-aware framework, we incorporate a modified focal frequency loss (FFL) that quantifies the distance between HR and SR OCT images in the frequency domain \cite{9710849}. The FFL is defined as:
\begin{equation}
FFL = \frac{1}{MN}\sum^{M-1}_{u=0}\sum^{N-1}_{v=0}w(u,v)|F_{SR}(u,v)-F_{HR}(u,v)|^2
\end{equation}
The $F_{SR}$ and $F_{HR}$ denote the frequency representation of SR and HR OCT images acquired by Discrete Fourier transform (DFT); the M and N represent the image size; the $w(u,v)$ is the spectrum weight matrix that is defined by:
\begin{equation}
w(u,v)=|F_{SR}(u,v)-F_{HR}(u,v)|^{\alpha}
\end{equation}
where $\alpha$ is the scaling factor for flexibility (set to 1 in our experiments).
In \cite{9710849}, the $F_{SR}$ and $F_{HR}$ are acquired by 2D DFT. However, the OCT images are acquired by 1D A-line scanning. Thus, we modify the FFL by acquiring $F_{SR}$ and $F_{HR}$ using 1D DFT. The original FFL is denoted as $FFL_{2D}$ and the modified FFL is denoted as $FFL_{1D}$. 
The loss function $L$ of the proposed frequency-aware model is defined as:
\begin{equation}
\begin{split}
&L(G, D, L1, FFL_{1D}, FFL_{2D}, L_{align}) = 
\\
& aL_{adv}(G, D) + bL1(SR, HR) + cFFL_{1D}(SR, HR) 
\\
&+dFFL_{2D}(SR, HR) + eL_{align}(G, D)
\end{split}
\end{equation}
The $L_{adv}$ stands for the adversarial loss; the $L1$ stands for the mean absolute error; the $L_{align}$ stands for the distance of high-frequency information between the HR and SR OCT images:
\begin{equation}
L_{align} = \sum^3_{i=1}|HF_{D}^i - HF_{G}^i|
\end{equation}
The factors $a$, $b$, $c$, $d$, and $e$ are set to 1, 10, 1, 1, and 1 respectively. We aim to solve the following minmax optimization problem:
\begin{equation} \label{eq2}
G^{*}=\arg \min\max L(G, D, L1, FFL_{1D}, FFL_{2D}, L_{align})
\end{equation}

\section{Results}
\subsection{Data collection}

We perform a large-scale quantitative analysis on an existing coronary image dataset. The dataset was imaged using a commercial OCT system (Thorlabs Ganymede, Newton, NJ) \cite{Li.2022}. The specimens were obtained in compliance with ethical guidelines and regulations set forth by Institutional Review Board (IRB), with de-identification to ensure the privacy of the subjects. A total of 2996 OCT images were obtained from 23 specimens, with a depth of 2.56 mm and a pixel size of 2 $\mu$m × 2 $\mu$m within a B-scan. The width of the images varied from 2 mm to 4 mm depending on the size of the specimen. 

In addition to the large-scale coronary, we also confirmed the generalizability of applying the proposed method to two small dataset: one from \textit{ex vivo} fish corneal and the other from \textit{in vivo} rat retina. Two fish corneal OCT images were obtained from the same Thorlab OCT images following the same imaging protocol as coronary imaging. Fifty rat retinal images were obtained from a Heidelberg Spectralis SDOCT system. The system has an axial resolution of 7 $\mu$m and a lateral resolution of 14 $\mu$m. The maximum field of view is 9 mm x 9mm. Animal imaging procedure was in accordance with protocols approved by the Institutional Animal Care and Use Committee at the Stevens Institute of Technology, and with the principles embodied in the statement on the use of animals in ophthalmic and vision research adopted by the Association for Research in Vision and Ophthalmology. The details of the experimental procedure are described in \cite{Liu.2020}.


\subsection{Experimental setup}
The OCT LR images from coronary images and eye images were generated by cropping the spectrum data. The optical resolution of OCT systems will be decreased by reducing the bandwidth of the spectrum. We used $\frac{1}{2}$,  $\frac{1}{3}$, and  $\frac{1}{4}$ (denoted as X2, X3, and X4 respectively) of the raw spectrum data by central cropping. We used Hanning window to filter the raw spectrum data. Next, the filtered spectrum data were processed by FFT to get complex OCT data. Then, the norm of the complex OCT data was converted to dB scale. The background subtraction was performed to remove noises in the OCT data. The LR OCT images were used as the inputs for the DL networks. The OCT images were randomly shuffled into five folds for cross-validation. 
\subsection{Network implementation}
We implemented our frequency-aware model using Pytorch. For downsampling layers, we used 2D convolutional layers with a stride of 2 followed by the Instance normalization layer and LeakyRelu activation layer with a negative slope of 0.2. For the upsampling layers, we used 2D transpose convolutional with a stride of 2 followed by the same Instance normalization and LeakyRelu activation. We used 16 residual blocks as the bottleneck, each containing 2 convolutional layers. For the implementation of previous DL algorithms (MSRN, RDN, RDU, RCAN), we were inspired by the designs in \cite{Li.2022}. We implemented the existing DL algorithms in GAN architecture.

\begin{figure*}[t]
\centering
\includegraphics[width=\textwidth]{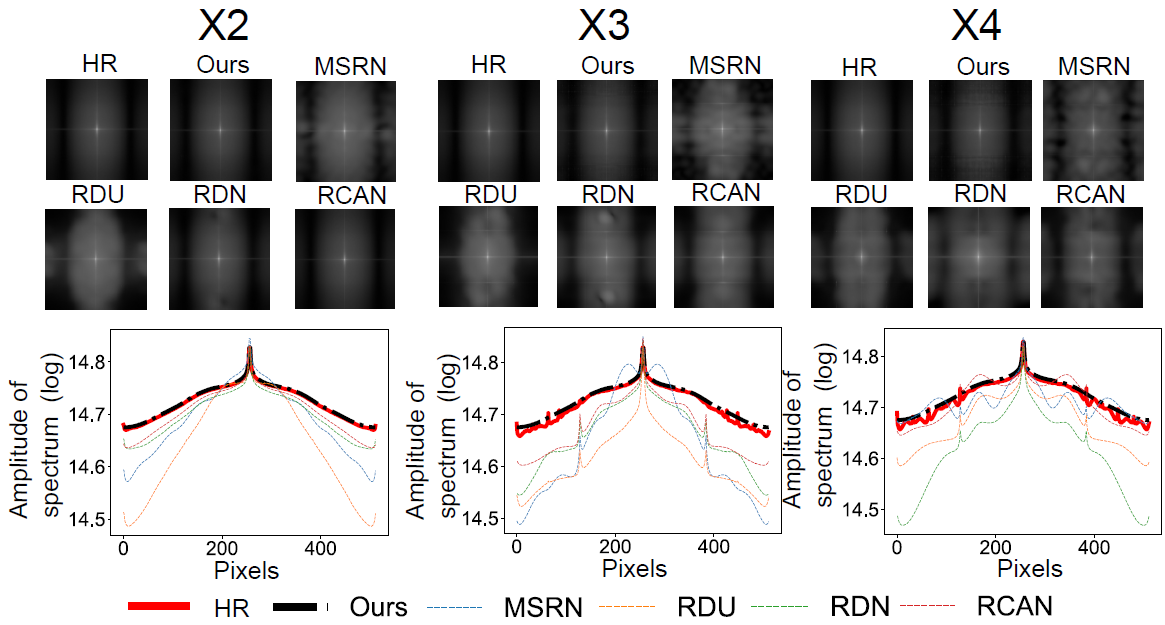}
\caption{Frequency analysis of the SR OCT images generated from LR OCT data acquired using factors of X2, X3, and X4. Compared to existing methods, our frequency-aware model is capable of super-resolving OCT images with less spectrum bias, which is confirmed by frequency analysis.}
\label{fig:lessBias}
\end{figure*}

\subsection{Training details}
The image intensities were normalized to a range of [0,1]. The training protocol was performed five times for cross-fold validation. We randomly sampled 16 non-overlapping LR patches of size 64 x 64 pixels as input during training. The normalized images were augmented through random flipping to prevent overfitting. The optimization routine was carried out using the Adam algorithm, with an initial learning rate of $10^{-4}$. A total of 200 epochs were executed to ensure convergence. The training process utilized one RTX A6000 GPU. Each training iteration on a single data fold consumed approximately 2 hours.
\subsection{Evaluation metrics}
We used peak signal-to-noise ratio (PSNR) and structural similarity (SSIM) \cite{1284395} to measure the quality of SR images. The PSNR calculates pixel-wise differences between the HR image and the SR image, which is defined as:
\begin{equation}
PSNR = 10\log_{10}(\frac{255^2}{MSE})
\end{equation}
The MSE represents the cumulative squared error between the HR and SR OCT images:
\begin{equation}
MSE = \frac{\sum^{M}_{m=1}\sum^{N}_{n=1}|HR(m,n)-SR(m,n)|^2}{M*N}
\end{equation}
The SSIM focuses on structural similarity between the HR image and the SR image, which is defined as:
\begin{equation}
SSIM = \frac{(2\mu_{HR}\mu_{SR}+c_1)(2\sigma_{HR,SR}+c_2)}{(\mu_{HR}^2+\mu_{SR}^2+c_1)(\sigma_{HR}^2+\sigma_{SR}^2+c_2)}
\end{equation}
where $\mu_{HR}$ is the pixel mean of the HR image; $\mu_{SR}$ is the pixel mean of the SR image; $\sigma_{HR}$ is the variance of HR; $\sigma_{SR}$ is the variance of SR; $\sigma_{HR,SR}$ is the covariance of HR and SR; $c_1=(0.01*255)^2$ and $c_2=(0.03*255)^2$ which are two variables to stabilize the division with weak denominator. 
\begin{table}[t]
\centering
\caption{PSNR, SSIM, and SFD results of OCT images reconstructed by MSRN, RDU, RDN, RCAN, and our frequency-aware model. \textcolor{red}{Red} indicates the best performance. All results are averaged based on five-fold cross-validation.}
\label{tableCompare}
\resizebox{0.9\columnwidth}{!}{%
\begin{tblr}{
  row{1} = {c},
  cell{1}{1} = {r=2}{},
  cell{1}{2} = {c=3}{},
  cell{1}{5} = {c=3}{},
  cell{1}{8} = {c=3}{},
  cell{3}{1} = {c},
  cell{4}{1} = {c},
  cell{5}{1} = {c},
  cell{6}{1} = {c},
  cell{7}{1} = {c},
  vline{3,6} = {1}{},
  vline{5,8} = {2-7}{},
  hline{1,3,8} = {-}{},
  hline{2} = {2-10}{},
}
\diagbox{Metrics}{Metrics}      & X2      &        &     & X3      &        &     & X4      &        &     \\
     & PSNR$\uparrow$    & SSIM$\uparrow$   & SFD$\downarrow$ & PSNR$\uparrow$    & SSIM$\uparrow$   & SFD$\downarrow$ & PSNR$\uparrow$    & SSIM$\uparrow$   & SFD$\downarrow$ \\
MSRN & 30.2094 & 0.8484 & 0.3765    & 24.2997 & 0.4829 & 0.8761    & 23.6034 & 0.4216 & 0.7565   \\
RDU  & 29.1010 & 0.7950 & 0.6737    & 23.8777 & 0.4628 & 0.8376    & 23.5086 & 0.4246 & 0.9369   \\
RDN  & 30.3321 & 0.8496 & 0.3503    & 23.7528 & 0.4635 & 0.6560    & 23.7746 & 0.4371 & 0.6977   \\
RCAN & 29.5280 & 0.8261 & 0.3918    & 23.5904 & 0.4596 & 0.7638    & 22.2740 & 0.3462 & 0.8121   \\
Ours & \textcolor{red}{30.4713}        & \textcolor{red}{0.8519}       & \textcolor{red}{0.3273}    &  \textcolor{red}{25.7500}       & \textcolor{red}{0.5904}       & \textcolor{red}{0.4669}    & \textcolor{red}{24.2867} & \textcolor{red}{0.4424} & \textcolor{red}{0.5287}     
\end{tblr}}
\end{table}
\begin{figure*}[t]
\centering
\includegraphics[width=\textwidth]{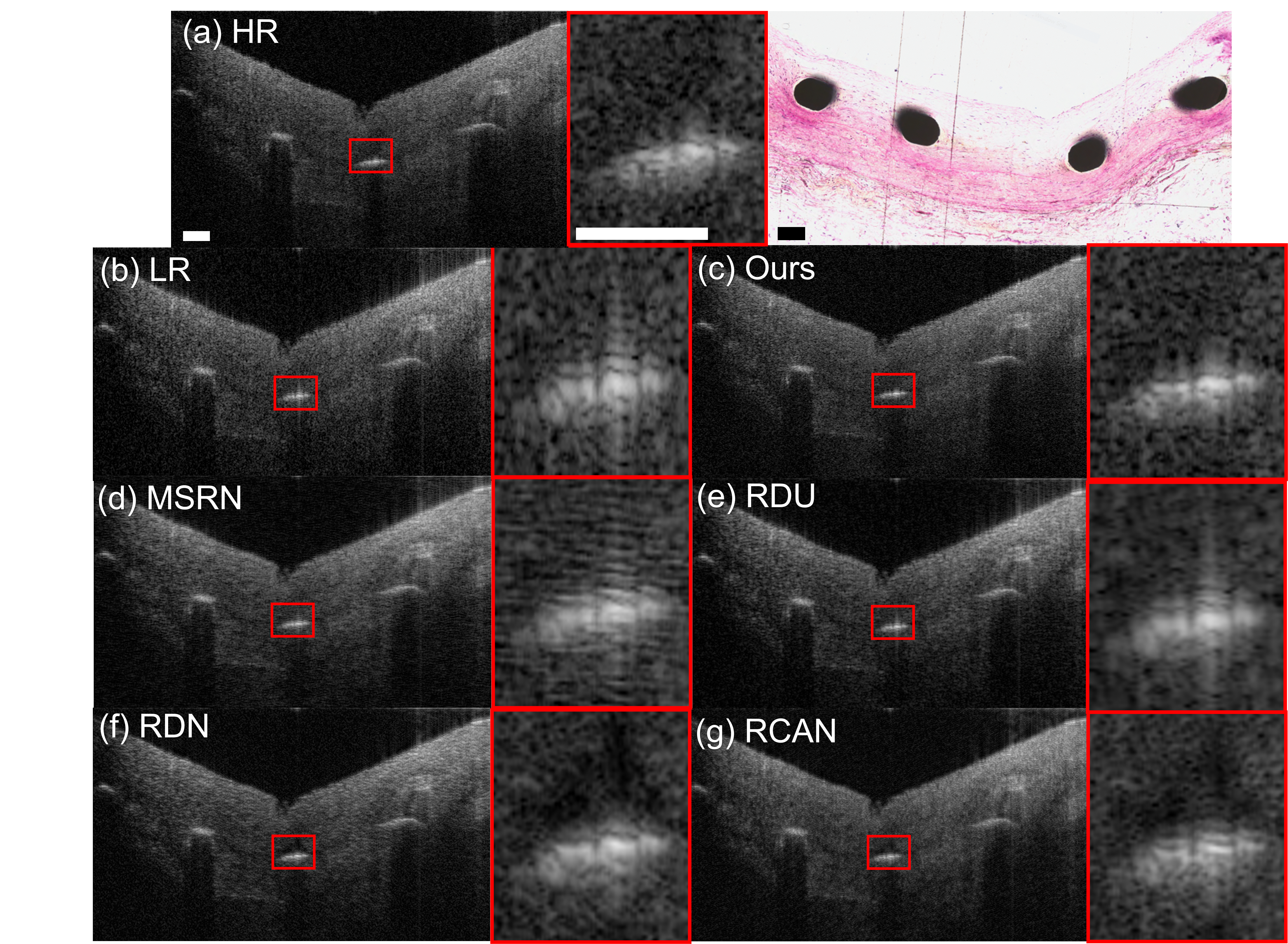}
\caption{Generating SR OCT images of stent structure from LR image acquired using a factor of X4. The corresponding histology image is attached. Our model resolves the boundary between the stent and tissue with better details due to its frequency-awareness design. ROIs are marked by red rectangles. The scale bar represents 100$\mu$m.}
\label{fig:VisInp}
\end{figure*}

To evaluate the frequency difference, we define a frequency-level metric, namely Scaled Frequency Distance (SFD), which is defined as:
\begin{equation}
SFD = \sum^{M-1}_{u}\sum^{N-1}_{v}\bigg|\frac{|F_{SR}(u,v)| - |F_{HR}(u,v)|}{|F_{HR}(u,v)|}\bigg|
\end{equation}
\subsection{Analysis on spectral bias}

\begin{figure*}[t]
\centering
\includegraphics[width=1\textwidth]{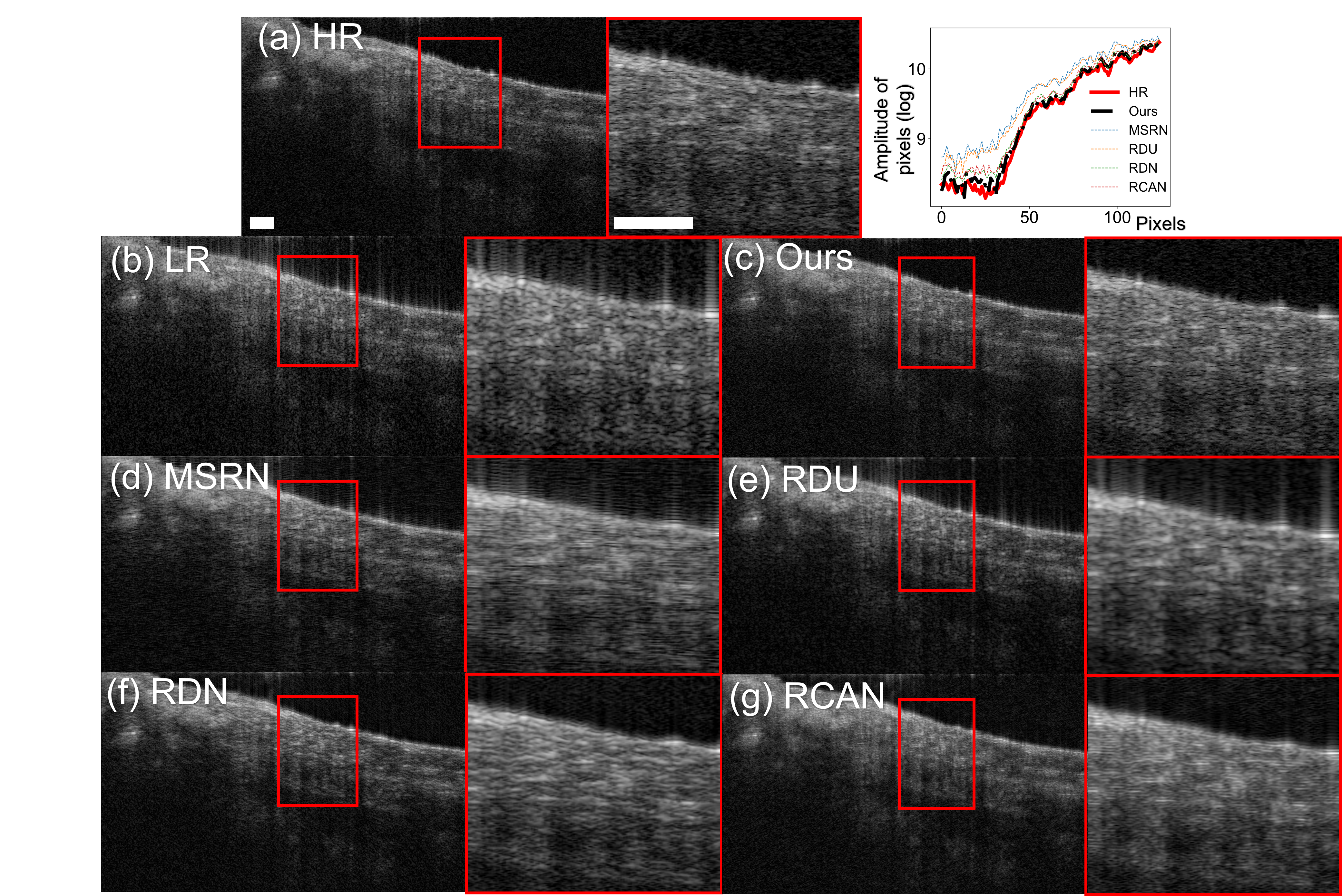}
\caption{Generating SR OCT image of suspicious macrophage regions from LR image acquired using a factor of X4. The amplitude of intensities of HR and SR images is attached. Our model resolves the accumulations of macrophages (indicated by the red arrows) without blurring effects. ROIs are marked by red rectangles. The scale bar represents 100$\mu$m.}
\label{fig:VisInpCoronary2}
\end{figure*}

We perform frequency analysis to evaluate the spectral bias of our frequency-aware model and other DL algorithms. We apply 2D DFT to the HR and SR OCT images, after which we average the logarithm of the intensities for each A-line and plot the intensity values over the pixels. The frequency analysis is carried out by averaging the spectrum of the SR OCT images. The results are reported in Fig \ref{fig:lessBias}. As shown in Fig \ref{fig:lessBias} (a), our frequency-aware model generates SR images with averaged spectrums that are similar to the HR images. The summed intensities for pixels, as shown in Fig \ref{fig:lessBias} (b), confirm our frequency-aware model are less biased in spectrum distribution compared with other DL algorithms. On the other hand, existing DL algorithms generate SR OCT images with spectral bias in an unstable manner, as confirmed by Fig \ref{fig:lessBias}.

\subsection{Quantitative analysis on super-resolution performance}

We compare the quantitative performance of our frequency-aware model to other DL algorithms. As shown in Table \ref{tableCompare}, our frequency-aware model generates SR OCT images with better PSNR, SSIM, and SFD scores compared to other deep learning algorithms. Together with Fig \ref{fig:lessBias}, we confirm our frequency-aware model generates SR OCT images with better spatial and frequency properties compared to other DL algorithms. 

In Fig \ref{fig:VisInp}, a case of super-resolving an LR OCT image of a stent within the coronary artery is demonstrated. Coronary stent placement is an established treatment for CAD \cite{ShpendElezi.1998}. Imaging microstructures and tissues adjacent to stent struts are crucial in the clinic. It is critical to provide accurate morphological information on interactions between the stent and the vessel wall, for the purpose of evaluating the placement as well as the biocompatibility of the stent. The edges of the stent are considered to be high-frequency information in the OCT images, which are challenging to reconstruct for previous DL algorithms. As shown in Fig \ref{fig:VisInp}, previous DL algorithms generate blurred edges of the stent. Moreover, existing DL algorithms lead to artifacts on the interaction between the stent and tissue, as shown in Fig \ref{fig:VisInp} (e), (f), (g). With our frequency-aware model, the edges of the stent are resolved with detailed information that is similar to that of the HR image. 

In Fig \ref{fig:VisInpCoronary2}, we demonstrate a case of super-resolving an LR OCT image of suspicious accumulation of macrophages. Macrophages play a critical role in both the development and rupture of atherosclerotic plaques \cite{TessaJ.Barrett.2020}, which are thus important for the diagnosis of CAD. OCT has been demonstrated to be a viable technique for visualizing the accumulation of macrophages in the human coronary artery. Macrophages appear as 'bright spots' in OCT images \cite{Tearney.2015}, which are high-frequency information due to their sharp contrast with neighboring tissues. As shown in Fig \ref{fig:VisInpCoronary2}, previous DL algorithms generate SR images with blurred macrophages, which will deteriorate the clinical diagnosis procedures. In contrast, our frequency-aware framework generates SR OCT images with clear macrophages.
Thus, our frequency-aware model is capable of providing SR OCT images for human coronary samples with clinical meaningness. 

\begin{figure*}[t]
\centering
\includegraphics[width=\textwidth]{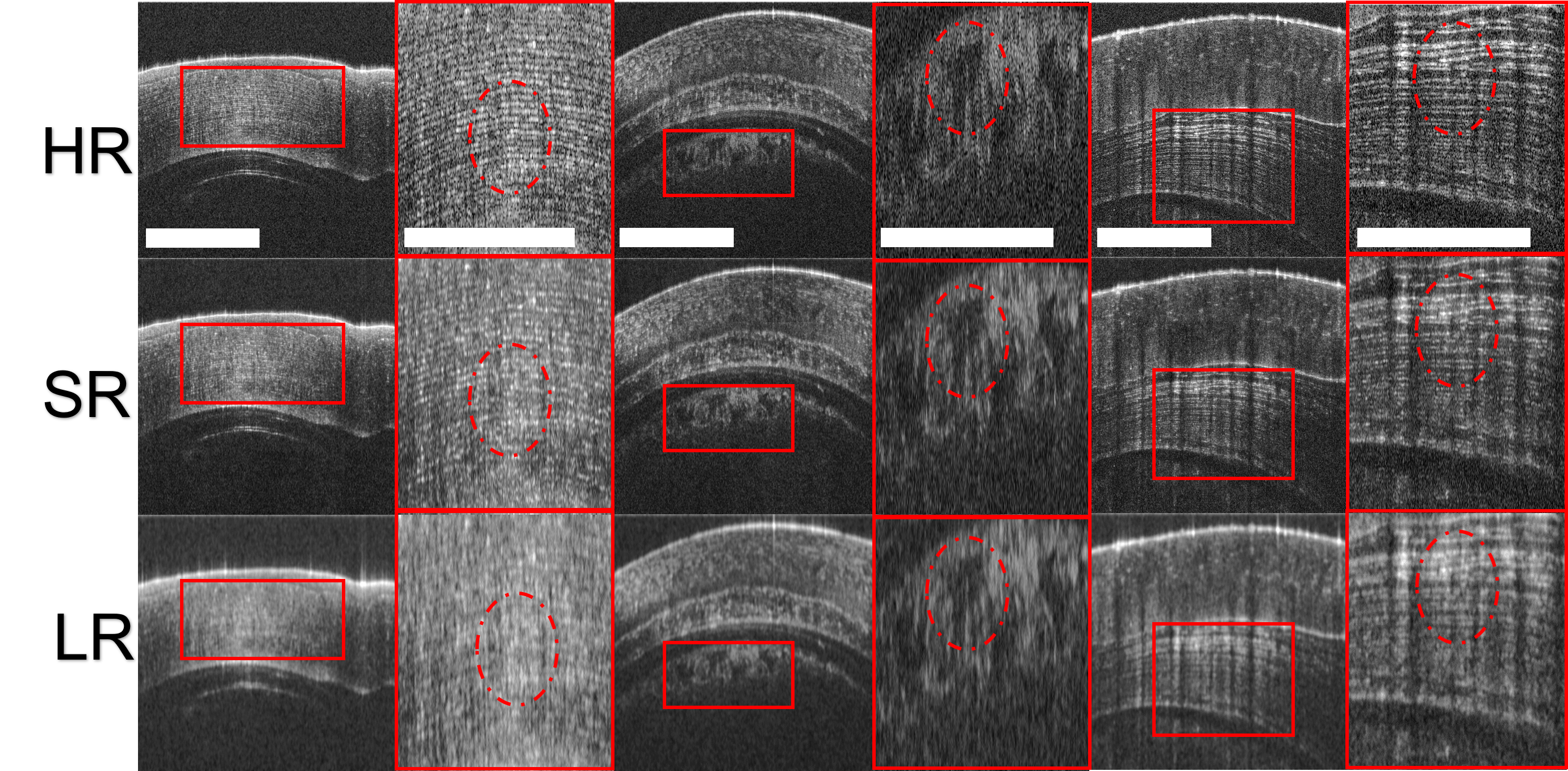}
\caption{Generating SR OCT images of anterior segments in fish eyes from LR images acquired using a factor of X3. ROIs are marked by red rectangles. The textures are highlighted by the dashed cycles. The scale bar represents 500$\mu$m.}
\label{fig:VisInp2}
\end{figure*}

\subsection{Application to super-resolve anterior segments of fish eye}
Based on the setup in coronary imaging, we perform additional experiments on fish cornea using trained frequency-aware framework from previous section. We acquired the fish corneal OCT images using the same OCT system and imaging settings as the coronary dataset. We acquired three volumes of left and right fish eyes. Three representative OCT B-scans are used for the qualitative studies. The qualitative analysis of SR OCT images of fish corneal is shown in Fig \ref{fig:VisInp2}. In particular, the dash circle in the first panel shows the alignment of corneal stroma is better resolved after super-resolving. The dash circle in the second panel highlights the iris region that is underneath the corneal regions. The dash circle in the third panel resolves the bownman's layer in corneal region. Overall, our frequency-aware framework is capable of generating SR OCT fish corneal images with sharper and finer textures. Without retraining, our frequency-aware framework has the potential to be transferrable to OCT corneal images obtained from the same OCT system.

\subsection{Application to super-resolve posterior segments of rat eye}
\begin{figure*}[t]
\centering
\includegraphics[width=\textwidth]{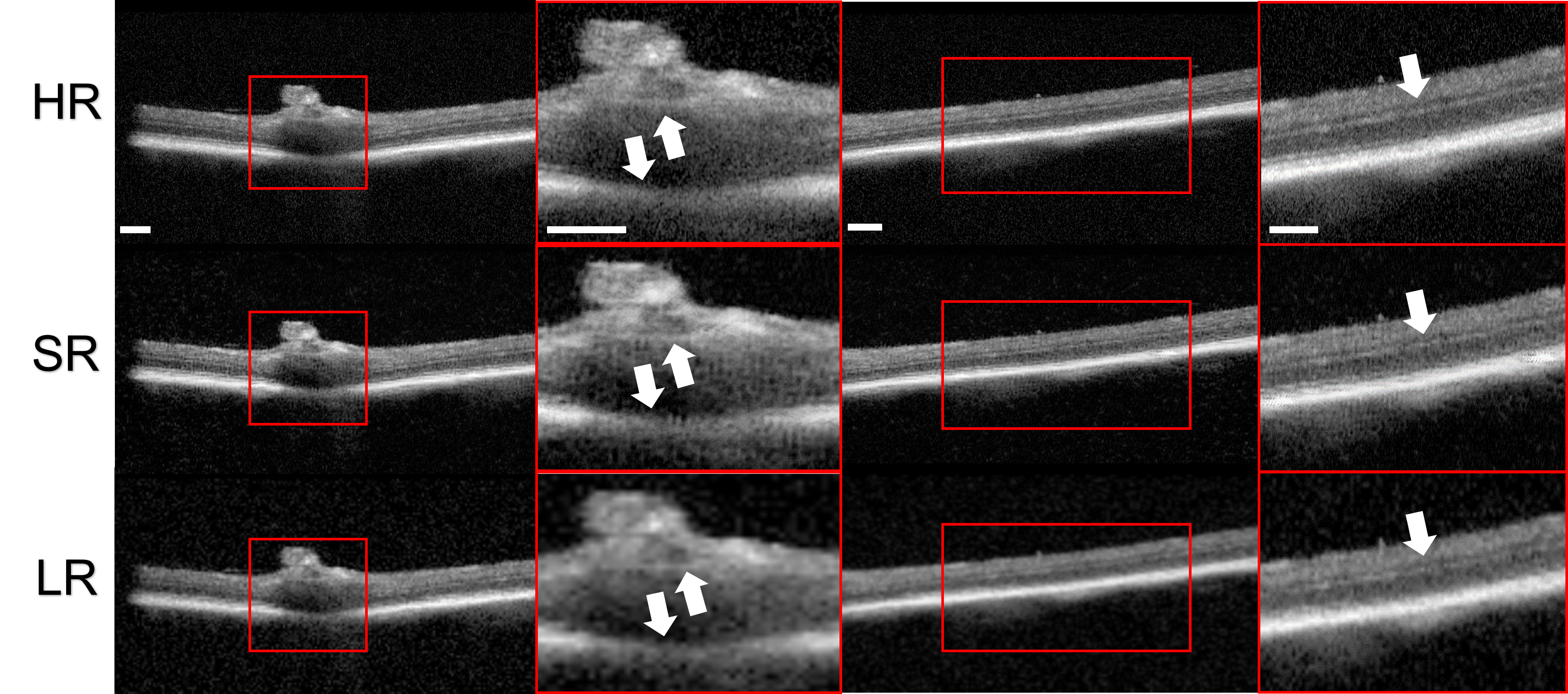}
\caption{Generating SR OCT images of posterior segments in rat eyes from LR images acquired using a factor of X3. ROIs are marked by red rectangles. The textures are highlighted by the white arrows. The scale bar represents 500$\mu$m.}
\label{fig:VisInp3}
\end{figure*}

We also conduct experiments to imaging posterior segments from animal model. A pigmented Long Evans rat from Charles River is used for OCT imaging. 
Retinal imaging requires a different optical design from benchtop OCT was used in coronary imaging and corneal imaging.
The OCT images were acquired using Heidelberg Spectralis system.  
We retrained the frequency-aware framework using $96\%$ of the images and used the rest of the images for testing. The SR OCT images of rat eyes generated by our frequency-aware framework are shown in Fig \ref{fig:VisInp3}. In the first panel, the SR OCT image delineates the boundary around optical disc; and in the second panel, the SR OCT image better resolves the layer boundary within retinal regions (for example, inner nuclear layers). 
This experiment shows that our proposed frequency-aware framework, with adequate retraining, has the potential to be generalizable to OCT retinal images obtained from different OCT systems.

\section{Discussion}
To the best of our knowledge, it is the first study in OCT community to propose a frequency-aware framework for super-resolution. We design the proposed framework by modifying the convolution model architecture and loss functions to improve the frequency-aware capability. This is based on our investigation on the spectral bias towards the low-frequency components in the spectrum in existing studies. Our frequency-aware model generates SR OCT images with less spectral bias and better performance compared to existing framework. Our frequency-aware model is capable of generating SR OCT images with clinical meaningness, which is confirmed by qualitative analysis. 

Another contribution lies in generalizability. Our preliminary study indicates great potential to be applied to multiple tissue types. We perform qualitative experiments on additional fish eye and rat eye dataset. Without retraining, our frequency-aware framework resolves anterior segments of fish eye, including corneal stroma, iris region, and downman's layer, acquired from the same OCT system. With adequate retraining, our frequency-aware framework is capable of resolving LR OCT images acquired from different systems. From the qualitative analysis of a rat eye dataset acquired from a different OCT system, we resolve the boundary around the optical disc and within retinal regions, with adequate retraining. 

As an exploratory study on methodology development, this study, especially the eye imaging, is based on healthy samples. In the future, we plan to validate our super-resolution framework on pathological animal models to examine how much the improved resolution could facilitate the diagnosis and treatment in ophthalmology. Moreover, in addition to SD-OCT, we plan to validate our super-resolution framework on swept-source OCT system in which the signal is also acquired in spectral domain. 

\section{Conclusion}
In this paper, we investigate the spectral bias of existing DL algorithms when generating SR OCT images. To mitigate the spectral bias, we develop a frequency-aware model that combines cGAN with frequency loss to super-resolve LR OCT images. 
Compared to existing DL algorithms, our approach produces SR OCT images with less spectrum bias, resulting in better preservation of textures. Additionally, our method generates SR coronary OCT images of superior quality, with higher PSNR and SSIM scores, as well as lower SFD scores. 
Our frequency-aware framework demonstrates the capability of generating SR OCT coronary images to provide better diagnosis and treatment. Moreover, our study also indicates the ability of the proposed framework to be generalized to corneal imaging and retinal imaging. 

\section{Funding}
This work was supported in part by National Science Foundation  (CRII-1948540 and CAREER-2239810, YG), New Jersey Health Foundation (YG), National Institute of Health (R01EY029298 and R01EY032222, JJKM).
\section{Disclosures}
The authors declare no conflicts of interest.
 
\section{Data availability statement}
Data underlying the results presented in this paper are not publicly available at this time but may be obtained from the authors upon reasonable request.

\section{Acknowledgement}
The authors would like to thank Chaimae Gouya, Mohammed Attia, and Aaron Shamouil for their assistance in OCT data acquisition.
\bibliography{sample}

\end{document}